\documentclass[aps,amssymb,amsmath,prd,reprint]{revtex4-2}

\usepackage{graphicx}
\usepackage{dcolumn}
\usepackage{bm}

\begin{document}


\title{Novel reaction force for ultra-relativistic dynamics of a classical point charge}
\author{P. Mart\'in-Luna}
\author{D. Gonz\'alez-Iglesias}
\author{B. Gimeno}
\author{D. Esperante}
\altaffiliation[Also at ]{Electronics Engineering Department, Universitat de Val\`encia, Avinguda de l'Universitat, 46100, Burjassot, Spain.}
\author{C. Blanch}
\author{N. Fuster-Mart\'inez}
\author{P. Martinez-Reviriego}
\author{J. Fuster}
\affiliation{%
 Instituto de F\'isica Corpuscular (IFIC), Universitat de Val\`encia - Consejo Superior de Investigaciones Cient\'ificas, Calle Catedr\'atico Jos\'e Beltr\'an, 2, 46980, Paterna, Val\`encia, Spain \\
}%

\date{\today}

\begin{abstract}
The problem of the electromagnetic radiation of an accelerated charged particle is one of the most controversial issues in Physics since the beginning of the last century, representing one of the most popular unsolved problems of the Modern Physics. Different equations of motion have been proposed throughout history for a point charge including the electromagnetic radiation emitted, but all these expressions show some limitations. An equation based on the principle of conservation of energy is proposed in this work for the ultra-relativistic motion. Different examples are analyzed showing that the energy lost by the charge agrees with the Li\'enard formula. This proposed equation has been compared with the Landau-Lifshitz equation obtaining a good agreement in the range of application of the Landau-Lifshitz formula. 

\end{abstract}

\maketitle


\section{\label{sec:intro}Introduction}

When a charged particle is accelerated it radiates electromagnetic energy according to the well-known Li\'enard formula \cite{Jackson1999_classical_electrodynamics}. The energy lost comes from the particle's kinetic energy and consequently a reaction force must act on the radiating particle for accounting such loss of kinetic energy. The equation of motion of a charged particle including the reaction force has been a subject of interest during the last 100 years and it has become one of the most popular unsolved problems of Modern Physics. The first formula of the  reaction force was the Abraham-Lorentz formula \cite{abraham1905theorie, lorentz1916theory}, which was extended to a Lorentz-covariant form by Dirac in 1938 \cite{Dirac1938_LAD_equation}. This expression is called the Lorentz-Abraham-Dirac (LAD) equation and suffers from some problems. Firstly, the LAD equation involves the temporal derivative of the acceleration and hence the initial position and
velocity of the particle does not determine the solution uniquely. For this reason, one extra initial condition is needed compared to the Newtonian mechanics. Secondly, if the external force is zero the LAD equation admits solutions that accelerate exponentially in addition to the trivial solution of uniform motion. These solutions are called runaway solutions and violate the principle of conservation of energy. The runaway solutions can be avoided by artificially introducing a preacceleration, but then the causality is violated. Furthermore, the LAD equation does not agree with the theory of Compton scattering except in the weak field case \cite{Hartemann2005_ComptonScattering_LAD}.

As a consequence of the inherent difficulties of the LAD equation, several alternative formulations have been proposed. Mo and Papas proposed in 1971 a new radiation reaction force formula that avoids the runaway solutions as well as the preacceleration phenomenon, assuming a reaction force proportional to the acceleration of the particle rather than the velocity \cite{MoPapas1971_EquationMotion_PhysRevD.4.3566}. Ford and O'Connell proposed in 1991 an alternative equation based on the use of the generalized quantum Langevin equation for an electron with a finite size \cite{Ford_OConnell1991}. Hartemann and Luhman \cite{hartemann1995classical}, Yaghjian \cite{yaghjian2010relativistic}, and Hammond \cite{hammond2010relativistic} have proposed other equations, but the most famous reaction force is the Landau-Lifshitz (LL) equation \cite{Landau1975} that can be obtained assuming that the radiation reaction force is much smaller than the external electromagnetic force. But the LL equation can also be obtained as a limit of the self-force modeling the electron as a sphere \cite{medina2006radiation, griffiths2010abraham}. Thus, the LL theory has been proposed as the correct equation of motion of a classical point charge \cite{rohrlich2001correct} and it has even been recently verified experimentally measuring the emission spectra of electrons and positrons penetrating into aligned single crystals \cite{nielsen2021experimentalLL}. However, the LL equation is not totally correct since it predicts no reaction force when the particle has a linear acceleration although in this case the radiation losses are practically negligible for typical electric fields \cite{Jackson1999_classical_electrodynamics}. Moreover, the LL equation fails for electromagnetic fields with abrupt changes, i.e. for high frequencies and/or high intensities \cite{hammond2008radiation_ABRUPT}.

The purpose of this work is the derivation of a new equation that overcomes these difficulties. The article is organized as follows. In Sec. \ref{Sec_2} the basic principles and the approximations to obtain the new proposed reaction force are discussed. In Sec. \ref{Sec_comparison} the proposed reaction force is solved for some simple physical examples comparing the results with the LL equation. We conclude the paper with final remarks in Sec. \ref{sec_conclusions}.


\section{The proposed reaction force} \label{Sec_2}
Radiation from charged particles becomes important for ultra-relativistic particles and its emission is concentrated in the direction of motion of the particle with a symmetric angular distribution around this direction \cite{Jackson1999_classical_electrodynamics}. Consequently, we will assume that the reaction force is antiparallel to the velocity $\bf{v}$, i.e.

\begin{equation}
\mathbf{F}_{\mathrm{R}}=-F_{\mathrm{R}} \hat{\mathbf{v},}
\end{equation}

\noindent with $\hat{\mathbf{v}}=\mathbf{v} /\|\mathbf{v}\|$, in order to satisfy in average the principle of conservation of energy and momentum when the emission is produced. Note that $\|\|$ indicates the euclidean norm of a vector on the 3-dimensional space $\mathbb{R}^3$. On the other hand, in the average distance, called mean free path \cite{Burkhardt2007_Radiation}

\begin{equation}
\lambda=\frac{2 \sqrt{3} \rho}{5 \alpha \gamma},
\end{equation}

\noindent is emitted a photon of mean energy

\begin{equation}
\langle E\rangle=\frac{8}{15 \sqrt{3}} E_{c},
\end{equation}

\noindent where $E_{\mathrm{c}}=\frac{3}{2} \hbar c \gamma^{3} / \rho$ is the critical energy, $\rho$ is the instantaneous radius of curvature of the trajectory, $\gamma$ is the Lorentz relativistic factor, $\hbar$ is the reduced Planck constant, $c$ is the speed of light in vacuum and $\alpha \approx \frac{1}{137}$ is the fine-structure constant. Thus, the principle of conservation of energy during a small displacement $\Delta x$ gives the relation $F_{\mathrm{R}} \Delta x=\frac{\Delta x}{\lambda}\langle E\rangle$, i.e.

\begin{equation}\label{ecu.FR}
F_{\mathrm{R}}=\frac{\langle E\rangle}{\lambda},
\end{equation}

\noindent which depends only on $\gamma$ and $\rho$.
The radiation emitted by an extremely relativistic charge is approximately the same as that of a particle moving instantaneously in a circular arc of radius of curvature \cite{Jackson1999_classical_electrodynamics}:
\begin{equation}
\rho=\frac{\|\mathbf{v}\|^{2}}{\|\dot{\mathbf{v}}_{\perp}\|} \approx \frac{c^{2}}{\|\dot{\mathbf{v}}_{\perp}\|},
\end{equation}

\noindent where $\|\dot{\mathbf{v}}_{\perp}\|$ is the magnitude of the perpendicular component with respect to $\mathbf{v}$ of the acceleration. Note that the dot indicates a time derivative. The equation of motion of the particle will be
\begin{equation}\label{ecu.F_particle}
\dot{\mathbf{p}}=m \dot{\gamma} \mathbf{v}+m \gamma \dot{\mathbf{v}}=\mathbf{F}_{\mathrm{ext}}+\mathbf{F}_{\mathrm{R}},
\end{equation}

\noindent where $\mathbf{p}$ is the relativistic linear momentum, $m$ is the rest mass of the particle and $\mathbf{F}_{\mathrm{ext}}$ is the external force applied to the charged particle, e.g. the Lorentz force. Therefore, projecting the equation of motion in the perpendicular direction, we obtain
\begin{equation}
\dot{\mathbf{v}}_{\perp}=\frac{\mathbf{F}_{\mathrm{ext}, \perp}}{m \gamma} .
\end{equation}

\noindent Hence, substituting in Eq. (\ref{ecu.FR}), the following reaction force is obtained
\begin{equation} \label{ecu.FR_perp}
\mathbf{F}_{\mathrm{R}}=-\frac{\tau_{m}}{m c} \beta^{-4} \gamma^{2}\|\mathbf{F}_{\text {ext }, \perp}\|^{2} \hat{\mathbf{v}} \approx-\frac{\tau_{m}}{m c} \gamma^{2}\|\mathbf{F}_{\text {ext }, \perp}\|^{2} \hat{\mathbf{v}},
\end{equation}

\noindent where the characteristic time $\tau_{m}=\frac{2}{3} \frac{1}{4 \pi \varepsilon_{0}} \frac{e^{2}}{m c^{3}}$ has been defined; $e$ is the elementary charge, $\varepsilon_{0}$ is the vacuum electric permittivity and $\beta=v/c$. This proposed reaction force will be a great approximation for ultra-relativistic particles. Note that if the acceleration is parallel to the velocity, the Eq. (\ref{ecu.FR_perp}) erroneously predicts a null reaction force as e.g. the Landau-Lifshitz force. Nevertheless, for a given magnitude of applied force, the radiation emitted due to a perpendicular acceleration is a factor $\gamma^{2}$ larger than with a parallel acceleration \cite{Jackson1999_classical_electrodynamics}. Consequently,  (\ref{ecu.FR_perp}) suggests the addition of an identical term for the parallel component of the force without the factor $\gamma^{2}$. Thus, the following final expression is obtained

\begin{equation}\label{ecu.FR_total}
\mathbf{F}_{\mathrm{R}}=-\frac{\tau_{m}}{m c}\left(\gamma^{2}\|\mathbf{F}_{\mathrm{ext}, \perp}\|^{2}+\|\mathbf{F}_{\mathrm{ext}, \|}\|^{2}\right) \hat{\mathbf{v}}.
\end{equation}

\noindent As a first check, we see that in absence of the external force the reaction force is zero. An alternative demonstration using four-vectors is included in the Appendix.

\section{SPECIAL CASES AND COMPARISON WITH THE LANDAU-LIFSHITZ EQUATION}\label{Sec_comparison}
Now we will solve the equation of motion (\ref{ecu.F_particle}) assuming the proposed reaction force (\ref{ecu.FR_total}) for an electron (with charge $q=-e$) in different basic physical situations comparing the results with those of the Landau-Lifshitz reaction force equation \cite{Landau1975}

\begin{equation}
\begin{aligned}
&F_{\mathrm{LL}}^{\mu}=\tau_{m}\left(\frac{q}{c} \frac{\partial F^{ \mu \nu}}{\partial x ^{\gamma}} u_{\nu} u^{\gamma}+\frac{q^{2}}{m c} F^{\mu \gamma} F_{\nu \gamma} u^{\nu}+\right.\\
&\left.\frac{q^{2}}{m c^{3}}\left(F_{\nu \gamma} u^{\gamma}\right)\left(F^{\nu \alpha} u_{\alpha}\right) u^{\mu}\right),
\end{aligned}
\end{equation}

\noindent where $F^{ \mu \nu}$ is the electromagnetic field tensor \cite{Jackson1999_classical_electrodynamics} and $u^{\mu}$ is the four-velocity; the metric tensor is \linebreak $g^{\mu \nu}=\mathrm{diag}(1,-1,-1,-1)$.
In the three-dimensional form it can be written for an electron as \cite{Bulanov2011_3D_LL_PhysRevE.84.056605}

\begin{equation} \label{ecu.LL3D}
\begin{aligned}
\mathbf{F}_{\mathbf{L L}} &=\tau_{m}\left\{e \gamma\left(\left[\frac{\partial}{\partial t}+(\mathbf{v} \cdot \nabla)\right] \mathbf{E}+\left\{\mathbf{v} \times\left[\frac{\partial}{\partial t}+(\mathbf{v} \cdot \nabla)\right]\mathbf{B}\right\} \right)\right.\\
&+\frac{e^{2}}{m c}\left(\mathbf{E} \times c \mathbf{B}+c[\mathbf{B} \times(\mathbf{B} \times \mathbf{v})]+\frac{1}{c} \mathbf{E}(\mathbf{v} \cdot \mathbf{E})\right) \\
&\left.-\frac{e^{2} \gamma^{2}}{m c^{2}} \mathbf{v}\left[(\mathbf{E}+\mathbf{v} \times \mathbf{B})^{2}-\frac{1}{c^{2}}(\mathbf{v} \cdot \mathbf{E})^{2}\right]\right\}
\end{aligned}
\end{equation}

\noindent where $\mathbf{E}$ and $\mathbf{B}$ are the external electric and magnetic field, respectively.
\subsection{Motion perpendicular to uniform magnetic field}
In this first example, we are going to assume a uniform magnetostatic field $\mathbf{B}=B_0\hat{\mathbf{z}}$ and motion in the $xy$-plane. Then, the proposed reaction force (\ref{ecu.FR_total}) and the LL force (\ref{ecu.LL3D}) are simply

\begin{equation} \label{Fr_magnetic_field}
\mathbf{F}_{\mathbf{R}}=\mathbf{F}_{\mathbf{L L}}=-\frac{\tau_{m} e^{2}}{m} \gamma^{2} B_{0}^{2} \mathbf{v},
\end{equation}

\noindent where $v\approx c$ has been assumed.
The solution of the motion for the general initial conditions $(x_{0},y_{0}), (v_{x0},v_{y0})$ is
\begin{equation} \label{ecu.motion_B_field}
\begin{gathered}
v_{x}(t)=e^{-\sigma \tau}\left[v_{x 0} \cos (\omega_{c} \tau)+v_{y 0} \sin (\omega_{c} \tau)\right], \\
v_{y}(t)=e^{-\sigma \tau}\left[-v_{x 0} \sin (\omega_{c} \tau)+v_{y 0} \cos (\omega_{c}\tau)\right], \\
x(t)=x_{0}+v_{x0}I_{1}(\tau) +v_{y0}I_{2}(\tau), \\
y(t)=y_{0}-v_{x0}I_{2}(\tau)+v_{y0}I_{1}(\tau),
\end{gathered}
\end{equation}

\noindent where we have defined the auxiliary integrals
\begin{equation}
\begin{gathered}
I_{1}(\tau)=\int_{0}^{\tau} \frac{e^{-\sigma \tau}}{\sqrt{1-\beta_{0}^2 e^{-2 \sigma \tau}}} \cos (\omega_{c} \tau) d \tau, \\
I_{2}(\tau)=\int_{0}^{\tau} \frac{e^{-\sigma \tau}}{\sqrt{1-\beta_{0}^2 e^{-2 \sigma \tau}}} \sin (\omega_{c} \tau) d \tau,
\end{gathered}
\end{equation}
\noindent the following quantities $\omega_{c}=eB_{0}/m, \sigma=\tau_{m}\omega_{c}^2$ and the proper time $\tau$ that can be expressed in the laboratory frame time $t$ as
\begin{equation}
\tau=\frac{1}{\sigma} \ln \left(\frac{\delta-e^{2 \sigma t}}{\delta-1}\right)-t
\end{equation}
\noindent with $\delta=\frac{1-\gamma_{0}}{1+\gamma_{0}}$. Note that $\beta_0=v_0/c$, $v_0=(v_{x 0}^2+v_{y 0}^2)^{1/2}$ and $\gamma_{0}=(1-\beta_{0}^2)^{-{1/2}}$.

Furthermore, we can calculate the energy lost per unit time $-\frac{dE}{dt}=-mc^2\frac{d\gamma}{dt}$, obtaining 
\begin{equation}
-\frac{dE}{dt}=\frac{\tau_{m}\gamma^{2}}{mc}e^{2} B_{0}^{2} v^{2}
\end{equation}

\noindent that agrees with the Li\'enard formula \cite{Jackson1999_classical_electrodynamics}

\begin{equation}\label{ecu.Lienard}
P=\frac{\tau_{m} \gamma^{2}}{m c}\left(\dot{\mathbf{p}}^{2}-\beta^{2} \dot{p}^{2}\right),
\end{equation}

\noindent if it is particularized for the considered uniform magnetic field.

\begin{figure}[h!]
\includegraphics[width=0.95\columnwidth]{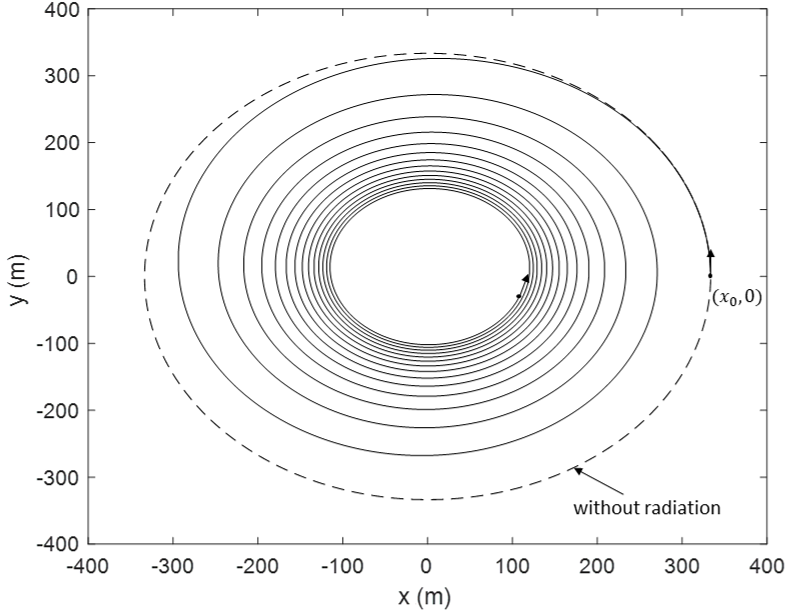}
\caption{Motion of an electron with an initial total energy of 100 GeV in a perpendicular magnetic field of $B=1$ T for the first 50 $\mu$s (the vectors indicate the initial and final velocity). The dashed line shows the circular trajectory with radius \linebreak $x_{0}=m \gamma_{0} v_{0} / (eB)$ obtained if the radiation is neglected.}
\label{fig:B_field}
\end{figure}

Figure \ref{fig:B_field} shows the trajectory (\ref{ecu.motion_B_field}) in the $xy$-plane of an ultra-relativistic electron. As it is expected, the motion is a spiral due to the radiated energy by the accelerated electron.

\subsection{Motion parallel to uniform electric field}
In this case we will assume a homogeneous static electric field $\mathbf{E}=E_0\hat{\mathbf{z}}$ and motion in the $z$-axis. Hence, the proposed reaction force (\ref{ecu.FR_total}) is

\begin{equation}
\mathbf{F}_{\mathbf{R}}=-\frac{\tau_{m} e^{2}}{m c} E_{0}^{2} \hat{\mathbf{v}}.
\end{equation}
and the Landau-Lifshitz force (\ref{ecu.LL3D}) is $\mathbf{F}_{\mathbf{LL}}=\mathbf{0}$. The general solution of this motion is

\begin{equation}
u_{z}(t)=\gamma(t) v_{z}(t)=u_{z 0}+\frac{q E_{0}}{m}(1-\mathrm{sgn}(v_{z})\epsilon) t,
\end{equation}

\begin{equation}
z(t)=z_{0}+\frac{m c^{2}}{q E_{0}(1-\mathrm{sgn}(v_{z})\epsilon)}(\gamma(t)-\gamma_{0}),
\end{equation}

\noindent where the dimensionless parameter $ \epsilon= \frac{\tau_{m} e E_{0}}{m c} \approx 3.68 \times 10^{-21} E_{0}(\mathrm{V/m}) $ has been defined and $\mathrm{sgn}(x)$ is the sign function. 
The solution of the Landau-Lifshitz force, i.e. neglecting the losses due to the radiation, is obtained by substituting $\epsilon=0$. Thus, it can be seen that for typical values of the electric fields in conventional particle accelerators, i.e, $E_{0}\sim 100 $ MV/m, this parameter includes a small perturbation to the motion obtained neglecting the effects of radiation. The cumulative energy lost as a function of time is given by
\begin{equation} \label{ecu.energy_E_field}
\Delta E(t)=m c^{2}(\gamma(t, \epsilon=0)-\gamma(t, \epsilon))=q E_{0} v_{z} t \epsilon+O\left(\epsilon^{2}\right).
\end{equation}

\noindent If we make the approximation $v_{z} \approx c$ this equation is analogous to the Li\'enard equation (\ref{ecu.Lienard}), as it is expected. Hence, our proposed reaction force will be valid as long as $\epsilon \ll 1$, i.e. up to electric fields of the order of $10^{19}$ V/m. It is important to remark that the effects of the radiation are in general totally negligible in this case. For example, an electron initially at rest accelerated with an electric field of \linebreak 1 GV/m during 100 m has finally an energy of \linebreak $\sim$100 GeV, but the predicted losses due to the radiation are $\sim$0.36 eV in such distance. However, Eq. (\ref{ecu.energy_E_field}) shows that the energy lost as a function of time is proportional to the square of the electric field. Consequently, if the electric field is increased four or five orders of magnitude (i.e. for electric fields $\sim$10-100 TV/m) the radiated energy will become very important. These ultrahigh electric fields have been predicted using X-ray wakefield acceleration in metallic crystals \cite{tajima1987crystal, ZhangTajima2016_AcceleratorCrystals_PhysRevAccelBeams.19.101004} and hence the LL equation cannot be used in their simulations, whereas the novel reaction force presented in this manuscript will be a good approximation.

\subsection{Motion in an electromagnetic plane wave}

In this section we are going to study the motion of an electron due to a linear polarized electromagnetic plane wave with fields $\mathbf{E}=E_{0} \cos (\omega t-k z) \hat{\mathbf{x}}$, $\mathbf{B}=B_{0} \cos (\omega t-k z) \hat{\mathbf{y}}$ with $\omega=ck$ and $E_{0}=c B_{0}$. The fourth order Runge-Kutta method has been used to numerically solve the electron relativistic dynamics. The radiated power can be calculated as 
\begin{equation} \label{ecu.power_radiated}
P_{\mathrm{rad}}=\mathbf{F}_{\mathbf{ext}}\cdot{\mathbf{v}}-\frac{dE}{dt},
\end{equation}
i.e., the difference between the power obtained from the external force (the Lorentz force) and the variation of the total energy per unit of time. 

\begin{figure}[!h]
\includegraphics[width=0.95\columnwidth]{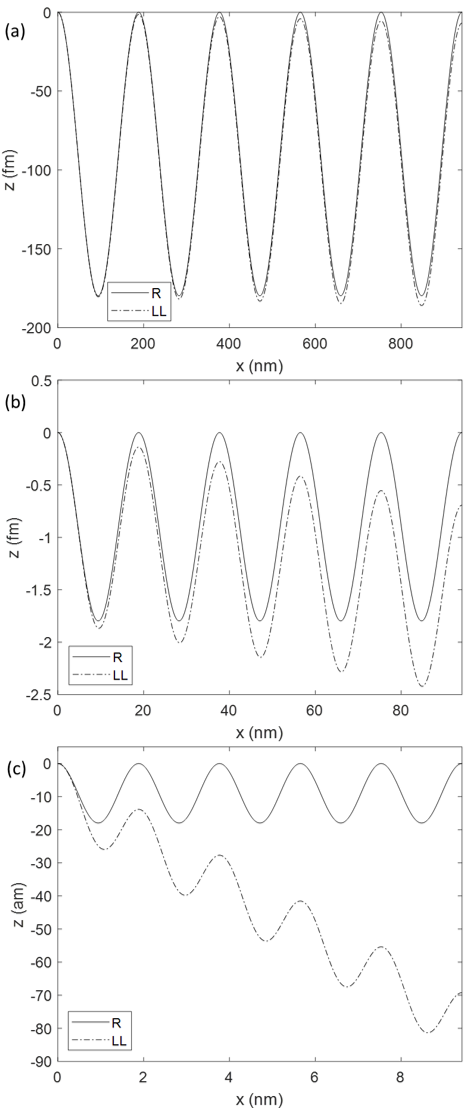}
\caption{Comparison between the trajectory obtained using the proposed reaction force (R) and the LL equation for $E_{0}=1$ TV/m and (a) $\omega=10^{16}$ rad/s, (b) $\omega=10^{17}$ rad/s, (c) $\omega=10^{18}$ rad/s. The initial conditions are $\mathbf{r_0}=\mathbf{0}$,  $\mathbf{v_0}=v_0\hat{\mathbf{x}}$ with an initial electron total energy of 10 GeV. }
\label{fig:plane_wave_trajectory}
\end{figure}

\begin{figure}[!h]
\includegraphics[width=0.95\columnwidth]{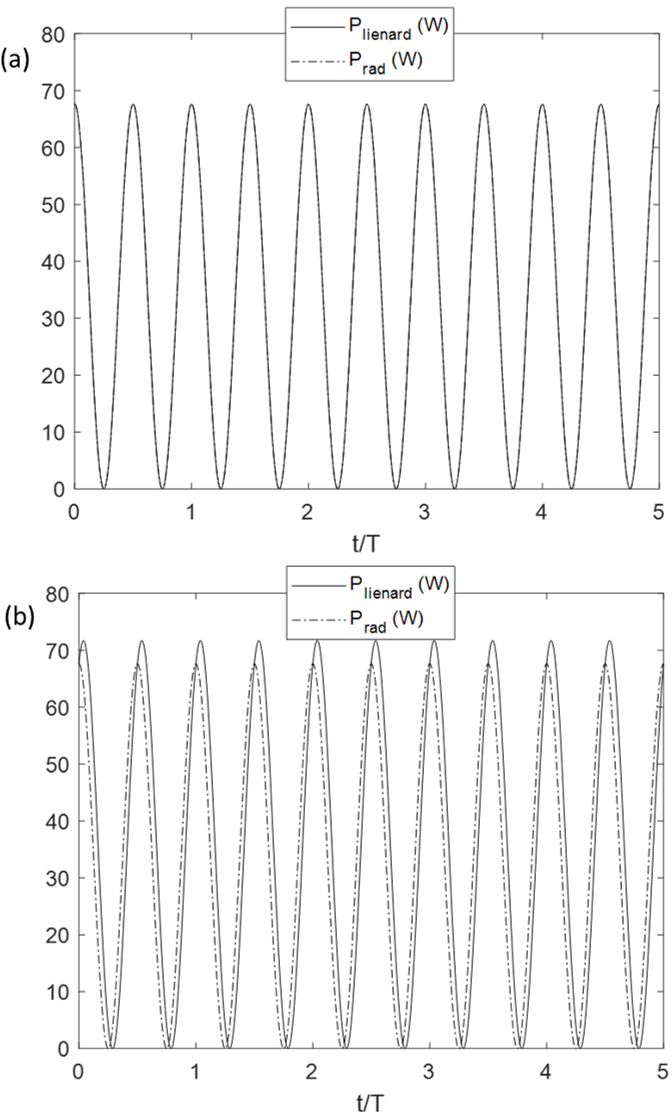}
\caption{Comparison between the Li\'enard formula and the radiated power associated with the motion obtained using the proposed reaction force (a) and the LL equation (b) for $E_{0}=1$ TV/m, $\omega=10^{18}$ rad/s, and the initial conditions $\mathbf{r_0}=\mathbf{0}$,  $\mathbf{v_0}=v_0\hat{\mathbf{x}}$ and an initial electron total energy of 10 GeV. The time is normalized to the period of the plane wave $T=2\pi/\omega$.}
\label{fig:plane_wave_radiated_power_comp}
\end{figure}

Figure \ref{fig:plane_wave_trajectory} shows the comparison between the trajectories obtained using the proposed reaction force (\ref{ecu.FR_total}) and the LL equation (\ref{ecu.LL3D}) for different frequencies $\omega$, showing that the disagreement between the trajectories becomes very important at higher frequencies. Furthermore, Fig. \ref{fig:plane_wave_radiated_power_comp} shows the comparison between the Li\'enard formula (\ref{ecu.Lienard}) and the radiated power (\ref{ecu.power_radiated}) when the electron relativistic dynamics is numerically solved using the proposed reaction force and the LL equation. It can be seen that the new reaction force agrees with the Li\'enard formula, while the LL equation does not have a good accordance. It is interesting to remark that the radiated power associated with the proposed reaction and the LL equation is similar, but the power predicted by the Li\'enard formula is different in each case since the motion is different (see Figure \ref{fig:plane_wave_trajectory}).

\begin{figure*}[ht]
\includegraphics[width=1.9\columnwidth]{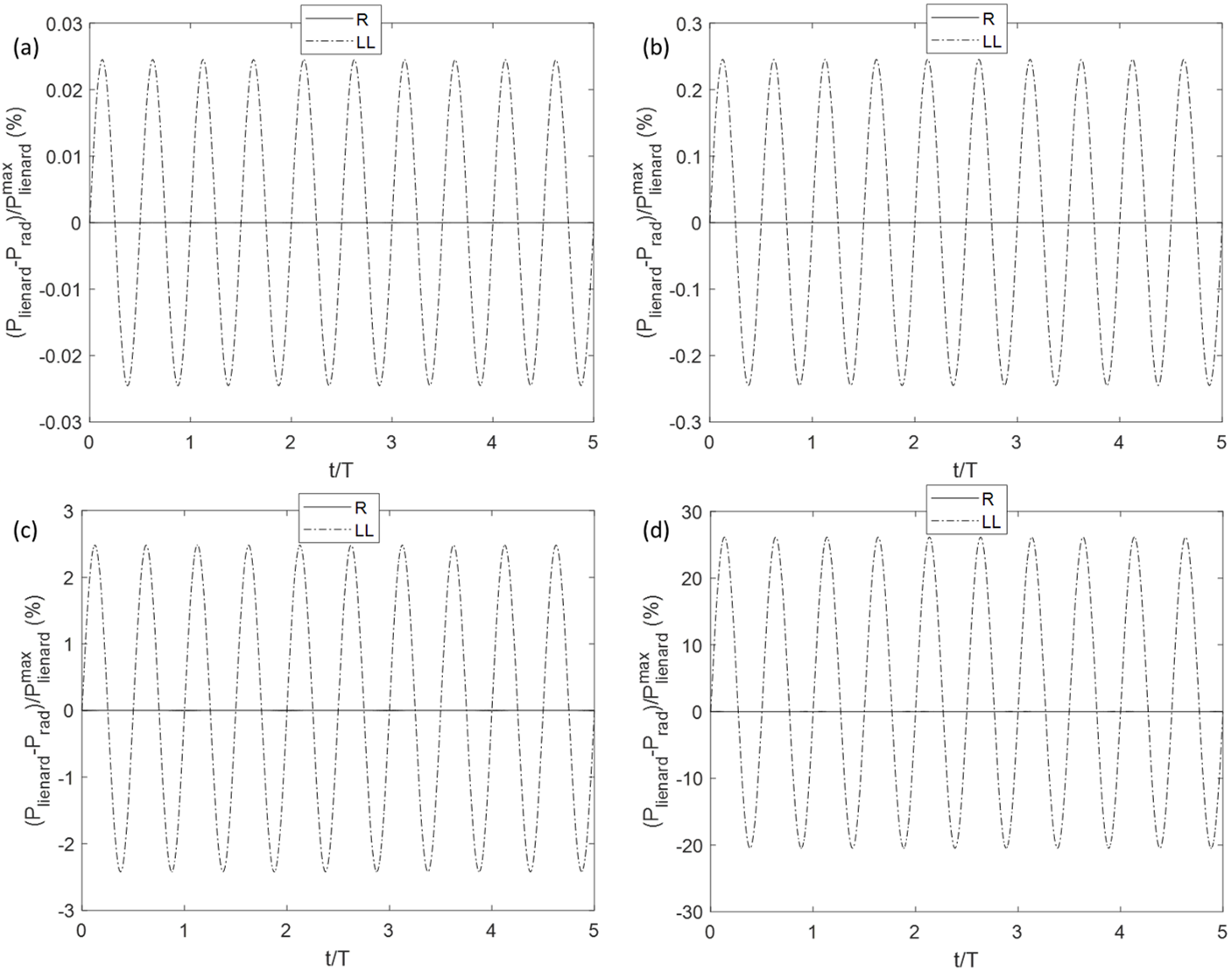}
\caption{Discrepancies between the Li\'enard formula and the radiated power associated with the motion obtained using the proposed reaction force (R) and the LL equation for different frequencies $\omega$: (a) $\omega=10^{15}$ rad/s, (b) $\omega=10^{16}$ rad/s, (c) $\omega=10^{17}$ rad/s and (d) $\omega=10^{18}$ rad/s. Same $E_{0}$ and initial conditions as in Fig. \ref{fig:plane_wave_radiated_power_comp}. The discrepancies are normalized to the maximum value of the corresponding Li\'enard formula and expressed in \%, and the time is normalized to the period of the plane wave $T=2\pi/\omega$. }
\label{fig:plane_wave_freqs}
\end{figure*}

Figure \ref{fig:plane_wave_freqs} shows the difference between the Li\'enard formula and the radiated power for different frequencies $\omega$ in order to study the differences between the two forces when the electromagnetic fields have very rapid changes. Firstly, we can see that the discrepancy between the radiated power obtained using the LL equation and the Li\'enard equation increases linearly with the frequency. Nevertheless, the discrepancy is negligible for the proposed reaction force. Therefore, it has been verified that the emitted radiation calculated with the proposed reaction force agrees much better to the Li\'enard formula for high frequencies than the LL equation which definitively fails.

\subsection{Motion in an undulator}\label{sec_undulator}

\begin{figure}[!h]
\includegraphics[width=0.95\columnwidth]{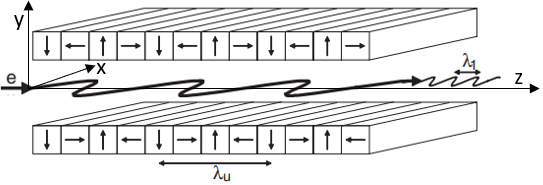}
\caption{Scheme of electron motion in an undulator. Adapted from \cite{Huang2007_FEL_review_PhysRevSTAB.10.034801}.}
\label{fig:undulator_scheme}
\end{figure}

Finally, we are going to consider an ultra-relativistic electron moving in a planar undulator as it is depicted in Figure \ref{fig:undulator_scheme}. In the planar undulator, near the $z$-axis, the magnetic field can be approximated by \cite{Huang2007_FEL_review_PhysRevSTAB.10.034801}

\begin{equation}
B_y(z)=B_0 \sin \left(\frac{2 \pi}{\lambda_u} z\right),
\end{equation}

\noindent where $\lambda_{u}$ is the undulator period length. If we consider a trajectory in the $xz$-plane and  assume that the energy is constant (i.e. $\gamma$ is constant), the horizontal component is given by

\begin{equation}\label{ecu.x_undulator}
x(z)=-\frac{K}{k_u \gamma \beta_z} \sin \left(k_{u}z\right),
\end{equation}

\noindent where we have imposed an initial velocity $v_{x}(z=0)=-Kc/\gamma, \|\mathbf{v}\|=\beta c$, $\mathbf{x_0}=\mathbf{0}$; $k_u=2 \pi / \lambda_u$, $K=\frac{e B_0}{m_e c k_u}$ is the undulator parameter and
\begin{equation}
\beta_z=\beta\left(1-\frac{K^2}{4 \gamma^2}-\frac{K^2}{4 \gamma^2} \cos \left(2 k_u z\right)\right)
\end{equation}
is the velocity (in units of $c$) of the electron in the $z$-direction.

In this example, the proposed reaction force is given by

\begin{equation} \label{Fr_undulator}
\mathbf{F}_{\mathbf{R}}=-\frac{\tau_{m} e^{2}}{m} \gamma^{2} B_{0}^{2}\sin^{2} \left(k_{u}z\right) \mathbf{v},
\end{equation}

\noindent and the LL equation is this Eq. (\ref{Fr_undulator}) plus a term due to the spatial derivative of the magnetic field. Nevertheless, for typical values of the undulator period (about tens of mm) this additional term is negligible since the magnetic field has not a sufficient strong spatial variation. Consequently, the discrepancies between the proposed reaction force and the LL equation when they are numerically solved are totally negligible. 

\begin{figure}[h!]
\includegraphics[width=0.95\columnwidth]{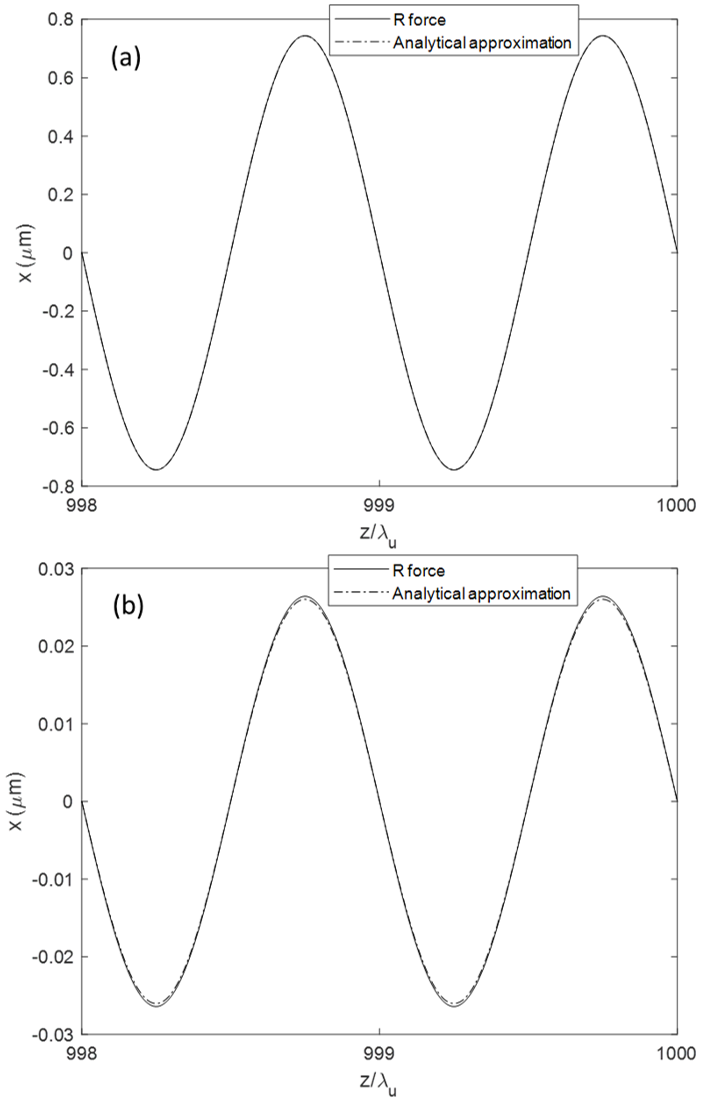}
\caption{(a) Comparison between the trajectory obtained using the proposed reaction force (R) and the analytical approximation (\ref{ecu.x_undulator}) of an electron with an initial total energy of  17.5 GeV and in a planar undulator with $K=4$ and $\lambda_{u}=40$ mm after 1000 periods. These numerical values correspond to the European XFEL parameters \cite{Smolyakov2014_undulators_spectra_IPAC2014}. (b) Same as (a) but with an initial energy of 500 GeV.}
\label{fig:undulator_trajectory}
\end{figure}

Figure \ref{fig:undulator_trajectory} displays the trajectory obtained solving the proposed reaction force compared to the Eq. (\ref{ecu.x_undulator}) showing that the difference is negligible for typical values even when the electron energy is 500 GeV. However, this approximation assumes no radiation and therefore cannot be used in order to take into account the effects of the electron electromagnetic energy loss in the radiation spectrum.


The energy lost per one undulator period can be calculated using the Li\'enard expression obtaining \cite{Smolyakov2014_undulators_spectra_IPAC2014}

\begin{equation}\label{ecu.undulator_energy_loss_period}
\Delta E(\text{period})=\frac{2}{3} \pi \alpha h c \frac{\gamma^2 K^2}{\lambda_u},
\end{equation}

\noindent and the relative energy loss $\delta$ is given by

\begin{equation}
\delta=\frac{\Delta E(\text{period})}{m \gamma c^2}=\frac{(2 \pi)^2}{3} \gamma K^2 \frac{r_e}{\lambda_u},
\end{equation}

\noindent where $r_e=\frac{\alpha \hbar}{m_e c}$ is the classical electron radius.

\begin{figure}[h!]
\includegraphics[width=0.95\columnwidth]{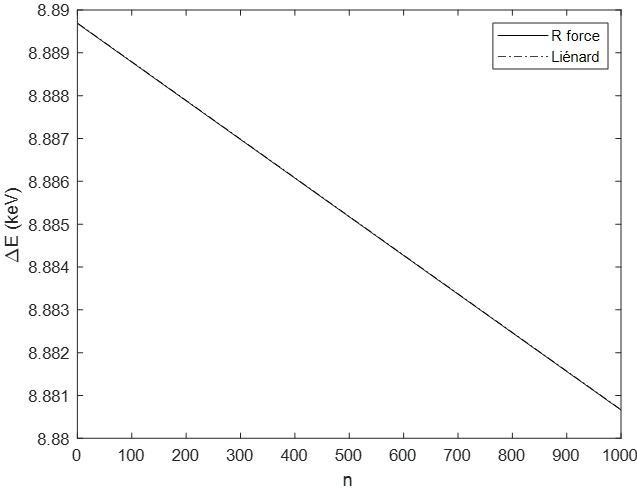}
\caption{Comparison between the energy lost per one undulator period using the proposed reaction force (R) and the Li\'enard expression (\ref{ecu.undulator_energy_loss_period}) as a function of the number of period $n$. The electron has an initial total energy of 17.5 GeV and the planar undulator parameters are $K=4$ and $\lambda_{u}=40$ mm. }
\label{fig:undulator_energy_loss}
\end{figure}

Figure \ref{fig:undulator_energy_loss} shows the energy lost per one period as a function of the number of period of the undulator. It can be seen that the result obtained with the proposed reaction force agrees with the Li\'enard expression. As a consequence of the electron energy 
decrease along the trajectory, the electromagnetic radiation from different undulators drops out of synchronism. Thus, the proposed reaction force can be easily used to calculate the radiation spectrum taking into account this effect. The interfering factor associated with the emission of electromagnetic radiation in each period of the undulator is given by \cite{Smolyakov2014_undulators_spectra_IPAC2014}

\begin{equation}
I_{\mathrm{int}}(\lambda)=\sum_{n=1}^N \exp \left(i \Phi\left(z_n\right)\right),
\end{equation}

\noindent where

\begin{equation}
\Phi\left(z_{n+1}\right)-\Phi\left(z_n\right)=2 \pi \frac{\lambda_{1, n}}{\lambda}
\end{equation}

\noindent is the difference of phase between the emission in the $(n+1)$-th and the $n$-th period of the undulator, $N$ is the total number of periods, $\lambda$ is the wavelength of the emitted radiation and

\begin{equation}
\lambda_{1, n}=\left(\frac{\widetilde{A C}}{\beta}-\overline{A C}\right)_n,
\end{equation}

\begin{figure}[h!]
\includegraphics[width=0.95\columnwidth]{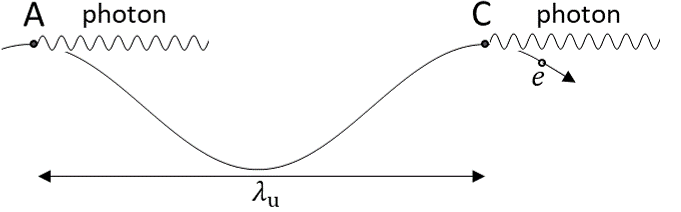}
\caption{Scheme of the electron trajectory in order to calculate the resonance condition of the emitted radiation.}
\label{fig:undulator_resonance}
\end{figure}

\noindent $\widetilde{A C}$ being the length of the electron trajectory and \linebreak $\overline{A C}=\lambda_{u}$ the length of the photon trajectory (see Figure \ref{fig:undulator_resonance}).

When we assume that $\gamma$ is constant, the interfering factor is given by
\begin{equation}\label{ecu.gamma_cte}
I_{\text {int }}(\lambda)=\frac{\sin \left(N \pi \frac{\lambda_1}{\lambda}\right)}{\sin \left(\pi \frac{\lambda_1}{\lambda}\right)}
\end{equation}

\noindent with $\lambda_1=\frac{\lambda_u}{2\gamma_{0}^2}\left(1+\frac{K^2}{2}\right)$. We can deduce from this expression that the resonance occurs at the wavelengths $\lambda=\lambda_{1}/k$ being $k$ an integer (the $k$-th harmonic) and the width of the maxima is inversely proportional to $N$. Otherwise, if we assume that the relative energy $\delta$ is constant, the phase is given by \cite{Smolyakov2014_undulators_spectra_IPAC2014}
\begin{equation}\label{ecu.delta_cte}
\Phi\left(z_n\right)=\frac{\pi \lambda_u}{\lambda \gamma_{0}^2}\left(1+\frac{K^2}{2}\right)(n-1)(1+\delta(n-2)),
\end{equation}

\noindent where $\gamma_{0}$ is the initial relativistic factor. On the other hand, using the proposed reaction force to numerically solve the electron dynamics, we can calculate the value of $\lambda_{1,n}$ as a discretization of the integral of the path described by the electron:

\begin{equation}\label{ecu.lambda_R}
\lambda_{1, n}=\left(\frac{\widetilde{A C}}{\beta}-\overline{A C}\right)_n=\left(\sum_{j}^{A \rightarrow C} \frac{|\Delta \vec{r}|_j}{\beta_j}-\left|z_C-z_A\right|\right)_n.
\end{equation}

\begin{figure}[h!]
\includegraphics[width=0.95\columnwidth]{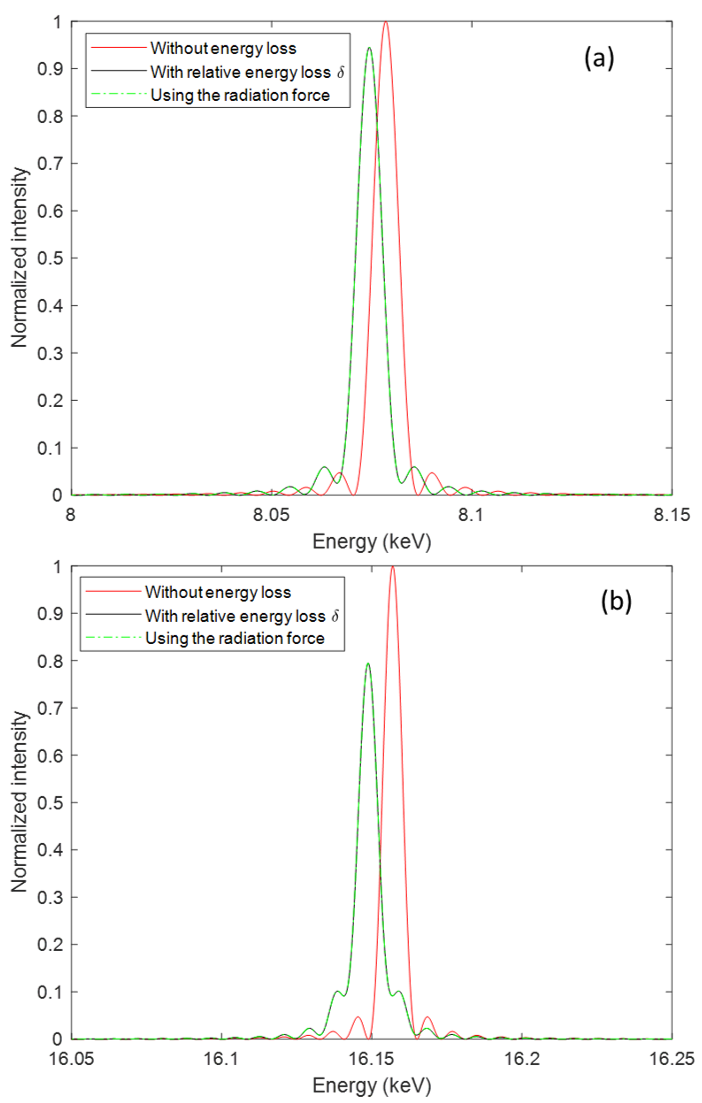}
\caption{Spectra of the electromagnetic radiation as a function of the photon energy $hc/\lambda$. Normalized intensity $|I_{\text {int }}/N|^2$ associated with an electron with an initial total energy of 17.5 GeV and a planar undulator with parameters: $K=4$ and $\lambda_{u}=40$ mm, $N=1000$. (a) First harmonic and (b) second harmonic.}
\label{fig:undulator_spectrum_1000_17p5}
\end{figure}

\begin{figure}[h!]
\includegraphics[width=0.95\columnwidth]{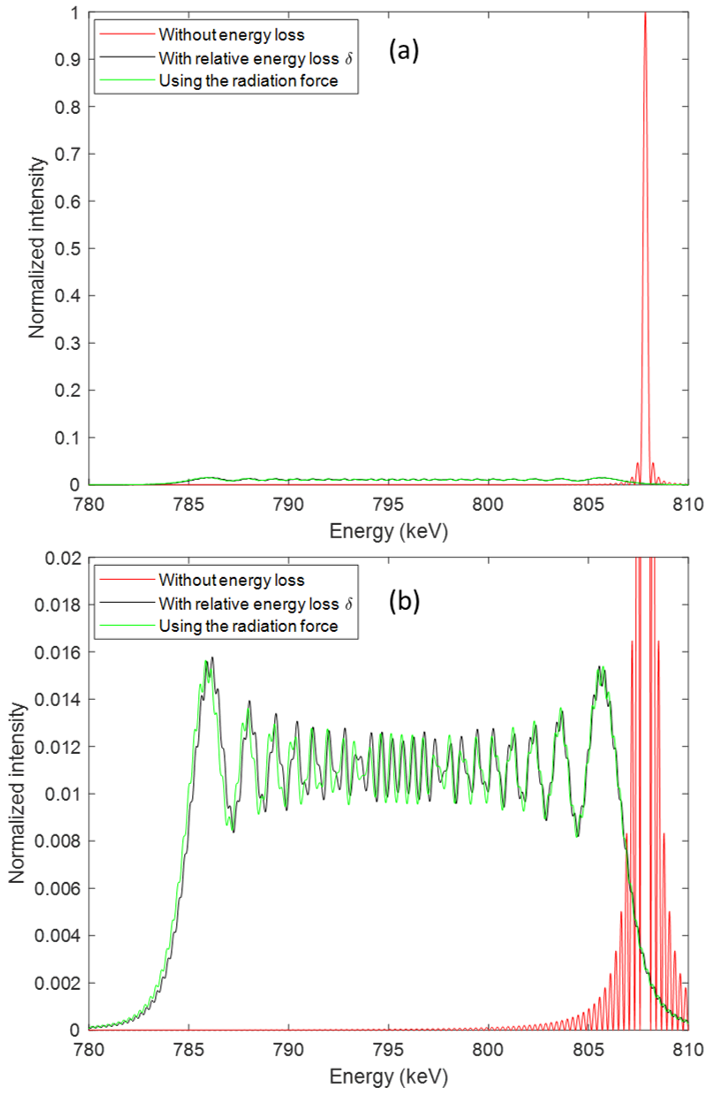}
\caption{Spectra of the electromagnetic radiation as a function of the photon energy $hc/\lambda$. (a) Normalized intensity $|I_{\text {int }}/N|^2$ associated with an electron with an initial total energy of 175 GeV and a planar undulator with parameters: $K=4$ and $\lambda_{u}=40$ mm, $N=3000$. (b) Zoom of the spectra calculated using Eq. (\ref{ecu.delta_cte}-\ref{ecu.lambda_R}).}
\label{fig:undulator_spectrum_3000_175}
\end{figure}

Thus, we are going to compare the interfering factor using the three different approximations: Eq. (\ref{ecu.gamma_cte})-(\ref{ecu.lambda_R}). Figure \ref{fig:undulator_spectrum_1000_17p5} shows the normalized intensity as a function of the energy of the emitted photons for the first and second harmonic. It can be seen that the approximation of $\delta$ constant (\ref{ecu.delta_cte}) and using the proposed reaction force (\ref{ecu.lambda_R}) agree very well. As a consequence of the energy lost by the electron along the trajectory, the peak maximum decreases and is redshifted. Furthermore, the spectrum is also wider compared to the case neglecting the energy loss. These three effects can be clearly seen in Figure \ref{fig:undulator_spectrum_3000_175} where the number of periods $N$ and the initial energy have been increased. In this case, the first harmonic has a very wide spectrum and the normalized intensity is lower than the 2 $\%$ of the peak maximum predicted neglecting the energy loss. Moreover, it can be seen that the approximation of $\delta$ constant agrees well with the case using the proposed reaction force, although in this extreme case there is a small discrepancy.

Thus, in this last example we have checked that the proposed reaction force is consistent with the typical approximations that are made in the undulators theory. Furthermore, the way that we have proposed in order to calculate the spectrum using the reaction force can be evidently used for any magnetic field. For example, the effects of the magnetic field nonuniformity along the horizontal X-axis due to finite poles could be easily studied. 

\section{Conclusions}\label{sec_conclusions}
A new reaction force has been proposed to take into account the electromagnetic radiation emitted from an accelerated charged particle. The proposed reaction force works equal or better than the Landau-Lifshitz force in the analyzed cases. Firstly, this new reaction force allows to introduce the radiation when the motion is parallel to an electric field, while the Landau-Lifshitz equation predicts no radiation in such case. Secondly, the proposed reaction force agrees much better than the Landau-Lifshitz force with the Li\'enard formula for electromagnetic fields with rapid changes. Moreover, this new reaction force is much easier to numerically solve and implement since it does not depend on the derivatives of the electromagnetic fields.

\begin{acknowledgments}
This work has been supported by Ministerio de Universidades (Gobierno de Espa\~{n}a) under grant number FPU20/04958.
\end{acknowledgments}

\section*{Appendix: Alternative demonstration of the proposed reaction force} \label{Appendix}

The Li\'enard formula for the radiated power in terms of the four-velocity is given by \cite{Jackson1999_classical_electrodynamics}

\begin{equation}
P_{\mathrm{rad}}=-\tau_{m}m\left(\frac{d u^{\mu}}{d \tau}\right)\left(\frac{d u_{\mu}}{d \tau}\right),
\end{equation}

\noindent where $\tau$ is the proper time. If we assume that the external force is the Lorentz force $F_{\mathrm{L}}^{\mu}=q F^{\mu \nu} u_{\nu}$, the equation of motion taking into account the radiation is

\begin{equation}
m \frac{d u^{\mu}}{d \tau}=F_{\mathrm{L}}^{\mu}+F_{\mathrm{rad}}^{\mu}.
\end{equation}

\noindent Thus, the temporal component is the variation of the total energy per unit of time

\begin{equation}
\frac{d E}{d t}=e \mathbf{E}\cdot{\mathbf{v}}+\frac{c F_{\mathrm{rad}}^{0}}{\gamma},
\end{equation}

\noindent where the second term is the contribution due to the radiated power $P_{\mathrm{rad}}$. Consequently, the temporal component of the reaction force is 

\begin{equation}
F_{\mathrm{rad}}^{0}=-\frac{\gamma P_{\mathrm{rad}}}{c}=-\frac{P_{\mathrm{rad}}}{c^{2}} u^{0},
\end{equation}

\noindent where the minus sign indicates that the particle is radiating energy. This equation suggests that the reaction force is proportional to the four-velocity, obtaining the following four-force

\begin{equation}\label{ecu.Frad4c}
F_{\mathrm{rad}}^{\mu}=-\frac{P_{\mathrm{rad}}}{c^{2}} u^{\mu}=\frac{\tau_{m}m}{c^{2}}\left(\frac{d u^{\nu}}{d \tau}\right)\left(\frac{d u_{\nu}}{d \tau}\right) u^{\mu}.
\end{equation}

\noindent If we assume that the reaction force is much less important than the Lorentz force, i.e. $m \frac{d u^{\mu}}{d \tau}=F_{\mathrm{L}}^{\mu}$, the vector component of (\ref{ecu.Frad4c}) is 

\begin{equation}\label{ecu.FR_total_appendix}
\mathbf{F}_{\mathrm{rad}}=-\frac{\tau_{m}v}{m c^{2}}\left(\gamma^{2}\|\mathbf{F}_{\mathrm{L}, \perp}\|^{2}+\|\mathbf{F}_{\mathrm{L}, \|}\|^{2}\right) \hat{\mathbf{v}},
\end{equation}

\noindent that coincides with the proposed reaction force (\ref{ecu.FR_total}) if $ v \approx c$ is assumed.

\providecommand{\noopsort}[1]{}\providecommand{\singleletter}[1]{#1}%
%


\end{document}